\documentclass[a4paper]{article}

\usepackage{atmohead2013}
\usepackage[english]{babel}

\title{A LIDAR system for the H.E.S.S. experiment}

\shorttitle{AtmoHEAD 2013 Template}

\authors{
M.Bourgeat,M.Compin,S.Rivoire,G.Vasileiadis
}

\afiliations{
LUPM,Un. Montpellier II
}

\email{george.vasileiadis@lupm.in2p3.fr}

\abstract{The H.E.S.S. experiment in Namibia, Africa, is designed to study the origin of high energy cosmic rays from 100Gev to few tens of TeV, using the Cherenkov technique. To miminize the systematic errors on the derived fluxes of the  measured sources, one has to calculate the impact of the atmospheric properties, namely the extinction parameter $\alpha$. A LIDAR can provide this kind of information within the detectable energy range of the experiment.  In this paper we report on the hardware components, operation and data taking of such a system installed on the HESS site for the last three years.}


\begin{document}
\maketitle

\section{Introduction}

The H.E.S.S experiment (for High Energy Stereoscopic System) consists of four imaging Cherenkov telescopes situated in the Namibia Khomas Highland desert (1800 m asl). Its main objective is the study of galactic or extragalactic sources in the energy  range of 100GeV to few tens of TeV coupled to a substantial flux sensitivity (1$\%$ Crab units). The combination of the four telescope data analysis provide good background rejection and angular resolution. The detection technique used by HESS, namely Cherenkov showers produced in the atmosphere by the incoming cosmic ray particle, demonstrate by itself the importance of knowing any variation on the atmospheric quality. 

Precise flux and energy spectra calculations for the observed sources could suffer if the level of cherenkov generated photons absorption due to aerosol or thin particles present in the atmosphere during the gamma shower development. Operating a Lidar during the data taking phase permit us to model the atmospheric transmission above the site, which in sequence could be used to simulate different background conditions and by varying the aerosol density, much the real background data \cite{bib:nol}. Once this achieved, these transmission tables are used to produce corrected tables for energy and effective area by by means of gamma-ray simulations.

\section{Atmospheric Monitoring}

The loss of Cherenkov light from a shower (as viewed by the H.E.S.S. telescopes) is mainly due to molecular and Mie scattering. The presence of aerosols can affect the data two-folded. If close to ground level the recorded photon yield can be lower affecting the telescope trigger thus affecting the reconstructed shower energy. If close to the shower maximum the shape and brightness of the camera images would be affected. The introduction of LIDAR measurements could permit to calculate the extinction parameter of the atmosphere $\alpha$ thus reducing the systematic error on the final energy spectrum calculation.
\begin{figure}
   \centering
   \includegraphics[width=8cm]{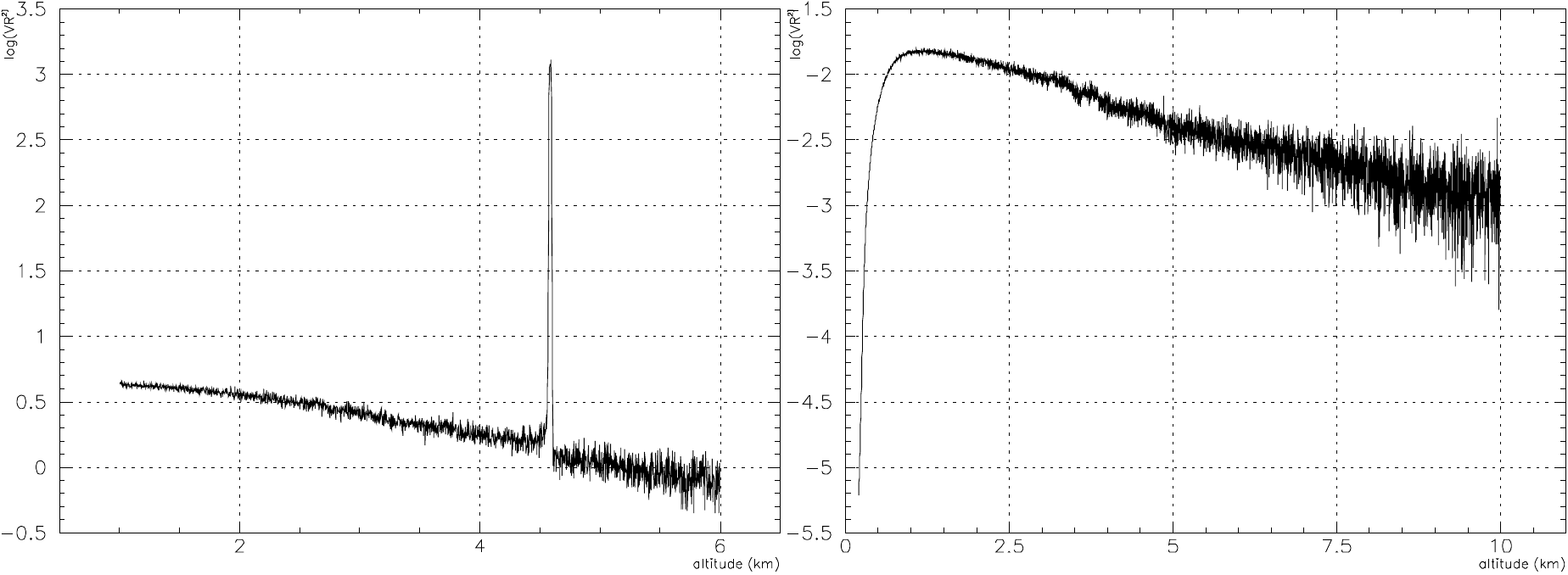}
      \caption{Profile of 1200 laser shots acquired during a period where at the beginning of the shift thin clouds were present, cleared later during the night. Both profiles are background subtracted and range corrected.}
         \label{fig1}
   \end{figure}
The back-scattered signal emitted by a LIDAR and observed from a PMT from a distance R is given by the LIDAR equation :
\begin{equation}
P(R)=K\frac{\beta(R)}{R^{2}}e^{-2\tau(R)}
\end{equation}
where $\beta(R)$ is the back-scattering coefficient and $\tau(R)$ is the integral of the extinction coefficient $\alpha(r)$ along the path. Both these quantities are sums of the aerosol and molecular contributions: $\alpha(R)=\alpha_{mol}(R)+\alpha_{aer}(R), \beta(R)=\beta_{mol}(R)+\beta_{aer}(R)$. The molecular part can be evaluated with standard models, knowing the pressure and temperature vs. height from atmospheric measurements. On the other-hand the extraction of the aerosol coefficients ($\alpha, \beta$), based on the LIDAR equation, demands algorithms based on either the Klett or Fernald methods. Since both methods need a priori assumptions on the molecular vs. aerosol parameter, the calculated value of $\alpha$ comes with a relative important systematic error.

\section{Lidar Hardware and Data Acquisition}
The LIDAR Telescope was conceived in 1997 and construction was completed in 1999. Until 2001 it was installed at the Themis site (Pyrenees,France) for the purpose of the Celeste experiment. Due to poor weather conditions on site it was used only for short intervals which led to frequent mechanical failures. 

During summer 2004 it was transferred and installed at the H.E.S.S. site, in Namibia (Africa). It is actually housed in a dedicated hut that is situated at a distance of 850m from the HESS facilities, minimizing the possibility that the cameras of the telescopes sees the lidar's laser beam. It started routine data taking operation during  summer of 2006. 

\subsection{Mirror and Mount}
For the collection of the backscattered light, the lidar uses a Cassegrain type telescope. The  primary parabolic mirror is of $\Phi=60cm$ diameter with a focal length of $\phi=102cm$, while the secondary has a diameter of 8cm and $\phi=10cm$. The mirrors was produced by Compact using BK7 glass coated with aluminum and a reflectivity of $80\%$ in the rage 300-600nm.The average spot size in the focus is 1.5mm FWHM. It is mounted in a fully steerable alt-azimuth frame equipped with DC servomotors with a maximum speed of $5^{o}$. The absolute pointing direction is close to $0.7^{o}$ accuracy. A three point mounting system allows for alignment and collinearity of the mirror and laser beam.
The whole apparatus is installed in a 5x5m hut equipped with a motorized roof, protecting it from rain and harsh condition when the lidar is not in operation. All motors can be controlled either locally or remotely via Ethernet connection.
\subsection{Laser}
The choice of the laser for our purpose is dictated by the following requirements: the wavelength of the laser has to match as much as possible the energy spectrum of the detected Cherenkov photons (300-650nm). Thus a double wavelength laser is needed giving us two reference points to compare with this spectrum. Repetition rate should be high enough to reduce collection time; the laser power should be adjustable to avoid interference with the Telescope optical system.
To meet these requirements a Quantel Brilliant 20 Nd:YAG laser was used. It is equipped with two cavities generating the 2nd and 3rd harmonics at 532 and 355nm. The repetition rate is 10Hz while the per-pulse energy is 180mJ and 65mJ respectively. The laser is mounted aside the Telescope structure. It is guided in a bi-axial configuration at a distance of 43cm from the optical axis of the telescope.

\subsection{Signal Detection and Digitization}
A pair of Photonis XP2012B photomultipliers are used for backscatter light detection. The return signal is split by means of a Dichroid filter, mounted at the focal point, separating the 532nm from the 355nm component. We use the photon counting method as a measurement which results in a saturation of the signal for the first few hundred meters. The photomultiplier gain reaches the value of $3x10^{6}$ at 1450V. The whole 51mm-diameter photocathode window is exposed to maximize light detection.

The output signals are fed through a pair of 1.5m long cable to a two-channel 12bit Compuscope Octopus CAGE 8265 Digitization board that runs at 65Mhz. This gives us an altitude resolution of 2.5m. A series of PCI boards commands the servomotors and hydraulic brakes. Finally a  Labview interface command the whole sequence of steering and data-taking of the Lidar system on a Windows XP based industrial CPU.
\subsection{Trigger and Operation Mode}
The laser system is been driven by means of a client-server protocol. The server is running on the H.E.S.S. DAQ system while the dedicated Lidar CPU run the client part. Upon reception of a acquisition demand, accompanied by several parameters as pointing direction, laser power and run duration, the Client execute the configuration asked, points the Telescope and triggers the laser system up to the nominal laser repetition rate. The DAQ is triggered by the laser synchronization pulse generated at every successful laser shot.

During normal H.E.S.S. operation the dedicated LIDAR runs are executed within the time interval that separates  the consecutive physics runs, 180sec interval. A 1000 laser shots profile  is executed  and stored in the HESS database in form of a ROOT file. 
\subsection{Signal Treatment}
The recorded signal is described by the so-called lidar equation :
\begin{equation}
P(r)=P_{0}\frac{ct_{0}}{2}\beta(r)\frac{A}{r^{2}}e^{-2\tau(r)}=P_{0}\frac{ct_{0}}{2}\beta(r)\frac{A}{r^{2}}e^{-2\int^{r}_{0}a(r')dr'}
\end{equation}
where $P(r)$ is the signal received at time $t$ from photons scattered at a distance $r$ from the lidar, $P_{0}$ is the transmitted laser power, $t_{0}$ is the laser pulse duration, $\beta(r)$ is the backscattering coefficient, $\tau(r)$ is the optical depth, $\alpha(r)$ is the extinction coefficient, and A is the effective area of the detector.  
 In cases where we can assume that the atmosphere is homogeneous, meaning there are no aerosols or clouds present, $\beta(r)=const.$ so the lidar equation can take the form:
\begin{equation} 
P(r)=P_{0}A\frac{ct_{0}}{2}\frac{\beta_c}{r^{2}}e^{-2\alpha(r)}
\end{equation}

The term $1/r^{2}$ in the lidar equation causes the measured signal $P(r)$ to diminished sharply with range because of the decreasing solid angle subtended by the receiving telescope with range. To compensate for this effect, we transform the receiving signal $P(r)$ into a range-corrected signal before the signal inversion begun. This is accomplished by multiplying the original signal $P(r)$ by the square of the range $r^{2}$. The range corrected signal denoted further $Y(r)$ can be written as :
\begin{equation}
Y(r)=P(r)r^{2}=P_{0}\beta_ce^{-2\alpha(r)}
\end{equation}
Taking the logarithm of the transform signal  and denoting it as $S(r)=lnY(r)$, one can rewrite the above equation as :
\begin{equation}
S(r)=ln(P_{0}\beta_c)-2\alpha(r)
\end{equation}
We see from the last formula that following  the assumptions mentioned above  the extinction coefficient has a linear dependency on $S(r)$ which is the issue behind the slope method for calculating the extinction parameter. Typical obtained results for a cloudy and clear night respectively are shown in Fig.~\ref{fig1} .
\section{Results and discussion}
\begin{figure}
   \centering
   \includegraphics[width=8cm]{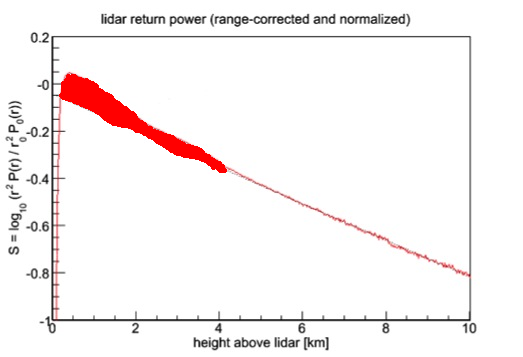}
      \caption{Qualitatively defined factor to characterize the presence of aerosol during data taking. Shaded expected to varies as a function of aerosol presence while its integral represents the level of the absorption.}
         \label{fig2}
   \end{figure}

During the last 2.5 years of operation we have collected 1670 atmospheric profiles combined with simultaneous data taking from the HESS experiment. The aim of the results that will be presented below is to achieve a correlation between the measured trigger rate of the HESS telescopes and any variation of the atmospheric absorption. As it is expected, in case of aerosol or cloud presence, the overall H.E.S.S. trigger rate will vary, since part of the cherenkov produced light will  be either absorbed or diffused. So one would expect that a strong correlation will evolve by looking at these two independently calculated quantities. In \cite{bib:tc} the method to calculate the HESS transparency coefficient (TC) is presented. This is designed as a hardware-independent as possible parameter  in order to separate hardware-related effects from the decrease in trigger rates caused by large-scale atmospheric absorption. It is thus much more preferable than the more standard global trigger rate.
From the LIDAR point of view, we define a parameter that qualitatively describes the aerosol presence as the ratio of the integral between the aerosol dominated region (usually between 1-4km) to a non aerosol one (6-9km) (shaded area in Fig.~\ref{fig2}). This parameter is expected to vary smoothly as the development of aerosol and their eventual extinction continues through the night, so as a first approximation, it was judged as the simplest but nevertheless accurate parameter to look for.

\begin{figure}
   \centering
   \includegraphics[width=8cm]{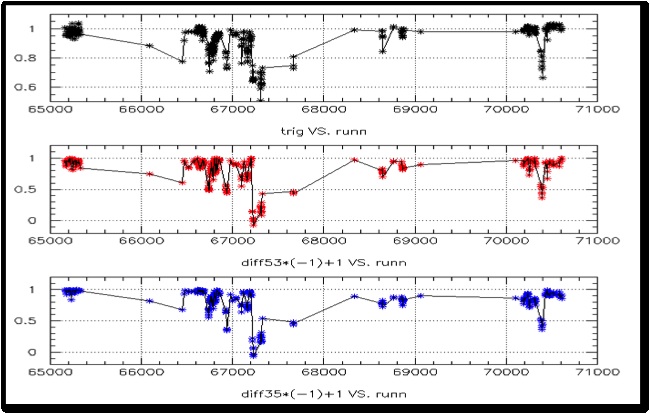}
      \caption{Timeline of the TC and aerosol presence for the last 2.5 years as measured on the H.E.S.S. site.}
         \label{fig3}
  \end{figure}

\begin{figure}
   \centering
   \includegraphics[width=8cm]{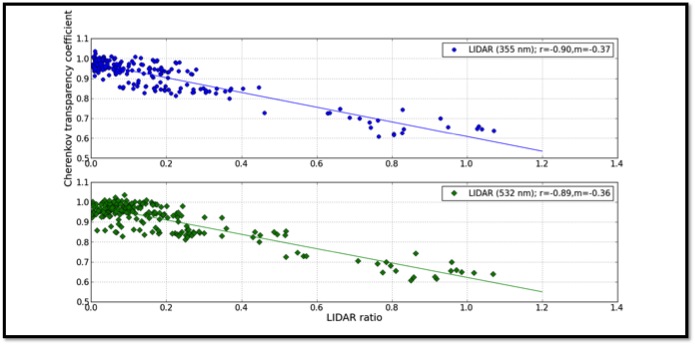}
      \caption{Scatter plot of the transparency factor and aerosol presence for the last 2.5 years of data taking on the H.E.S.S. site.}
         \label{fig4}
  \end{figure}

After closely selecting all runs that exclude cloud presence, since H.E.S.S. do not take data during these periods, we defined a list of atmospheric profiles to compare with the H.E.S.S. TC parameter mentioned above. Fig.~\ref{fig3} shows a timeline of these selected runs, where we clearly identify the similarities on the variation of TC compared to the aerosol presence on both LIDAR wavelength bands.  For simplicity we have inverted the aerosol absorption parameter to reflect the same trend as the TC one.  
Fig.~\ref{fig4} shows the same parameters in a scatter plot way where we clearly demonstrate the strong correlation between the aerosol presence and TC variation,  for both LIDAR detected wavelengths (355nm and 532nm).
\section{Conclusions}
We presented here, for the first time, a strong  correlation between the aerosol layer measured on the H.E.S.S. site for the last 2.5 years and the transparency factor, a quality assurance factor that takes into account all possible variations and impact on data quality apart the atmospheric one. A very clear influence on data rates and absorption of cherenkov light is evident whenever a aerosol layer is detected and measured by the H.E.S.S. LIDAR. Next steps on this analysis will include a more proper calculation of the aerosol layer using more appropriate methods (Klett\cite{bib:klett} ). Eventually the LIDAR based aerosol calculated absorption will be included in the overall TC factor calculated and used by the H.E.S.S. experiment.

\end{document}